\documentstyle[aps,prl,multicol,floats,epsf]{revtex}

\begin{document}

\draft

\wideabs{
\title{Magnetoresistance Anomalies in Antiferromagnetic
YBa$_2$Cu$_3$O$_{6+x}$: Fingerprints of Charged Stripes}

\author{Yoichi Ando, A. N. Lavrov\cite{ANL}, and Kouji Segawa}

\address{Central Research Institute of Electric Power Industry,
2-11-1 Iwato-kita, Komae, Tokyo 201-8511, Japan}

\date{stripe5fa.tex}

\maketitle

\begin{abstract}
We report novel features in the in-plane magnetoresistance (MR)
of heavily underdoped YBa$_2$Cu$_3$O$_{6+x}$, 
which unveil a developed ``charged stripe'' structure in this system.
One of the striking features is an anisotropy of the MR with 
a ``$d$-wave" like symmetry upon rotating the magnetic field $H$ 
within the $ab$ plane, which is caused by the rotation of the 
stripes with the external field.
With decreasing temperature, a hysteresis shows up below $\sim$20 K 
in the MR curve as a function of $H$ and finally below 10 K the 
magnetic-field application produces a persistent change 
in the resistivity.  This ``memory effect" is caused by the freezing 
of the directionally-ordered stripes.
\end{abstract}

}
\narrowtext

High-$T_c$ cuprates can be viewed as doped antiferromagnetic (AF)
insulators with the charge carriers transferred into perfect CuO$_2$ planes
from the charge reservoir layers. In general, there is a tendency for AF
insulators to expel doped holes and to show a phase separation into regions
with and without the doped holes \cite{Emery}. When the ions are not
mobil and the Coulomb interaction between doped holes is effective, 
an intriguing microscopic state with holes gathered within an array of 
quasi-1D ``stripes'' may be realized \cite{Emery,AFstrip1,AFstrip2}. 
An ordered striped structure has actually
been observed by the neutron diffraction in a doped antiferromagnet,
La$_2$NiO$_{4.125}$ \cite{Ni}. Inelastic neutron scattering studies of
superconducting (SC) cuprates found incommensurate magnetic fluctuations
\cite{incomm}, which can be considered as dynamical stripe correlations
\cite{Emery}; it has been discussed that such dynamical stripes are pinned
in La$_{1.6-x}$Nd$_{0.4}$Sr$_x$CuO$_4$, which shows a static striped
structure similar to that of the nickelate except for the direction of the
stripes \cite{LaNd}.

Whether the microscopic charge inhomogeneities in the form of dynamic or
static stripes, if present at all, are relevant to the unusual normal-state
and SC properties in cuprates remains unclear.  For SC compositions, the
fast dynamics of the stripe correlations \cite{incomm} make it implausible
to study them by methods other than spectroscopy.  In fact, very little is
known about the electron dynamics in the stripes.  
The existence of the static stripes 
in cuprates has been documented at least for two cases: (a) in
La$_{1.6-x}$Nd$_{0.4}$Sr$_x$CuO$_4$ \cite{LaNd}, where the stripes are
pinned by lattice deformations, and (b) in heavily-underdoped,
non-superconducting cuprates \cite{AFstrip1}, in which the dynamical stripe
correlations presumably crossover into a static stripe array as AF domains
grow in the CuO$_2$ planes. To study the detailed properties of the
stripes, it is natural to turn to either of these systems.

In this Letter, we present novel magnetoresistance (MR) features observed
in heavily underdoped, antiferromagnetic YBa$_2$Cu$_3$O$_{6+x}$, which can
hardly be explained without assuming a hole segregation into some flexible
stripes. The most intriguing observation is the possibility to induce a
persistent in-plane anisotropy just by exposing a crystal to the magnetic
field, upon which the higher conductivity axis is adjusted to the field
direction. This phenomenon can be understood to originate from a
field-induced topological ordering of the stripes, whose direction remains
unchanged after removing the magnetic field at low temperatures.

\begin{figure}[t]
\epsfxsize=8.5cm
\centerline{\epsffile{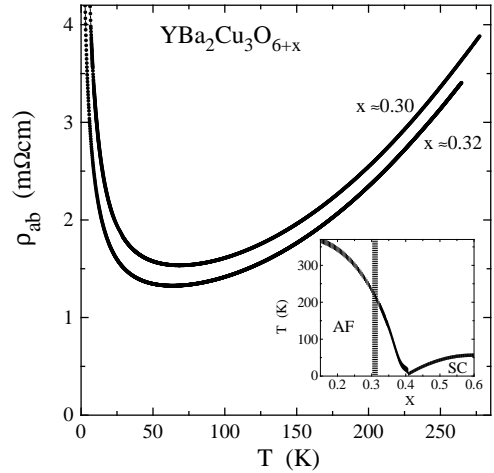}}
\vspace{2mm}
\caption{$\rho_{ab}(T)$ of the two YBa$_2$Cu$_3$O$_{6+x}$ single
crystals. Inset illustrates the part of the phase diagram being studied.}
\label{fig1}
\end{figure}

The high-quality YBa$_2$Cu$_3$O$_{6+x}$ single crystals were grown by the
flux method in Y$_2$O$_3$ crucibles, and a high-temperature
annealing was used to reduce their oxygen content. The exact oxygen 
content was determined by iodometry \cite{Segawa}.
The magnetoresistance
was measured using a standard ac four-probe technique by sweeping the
magnetic field at fixed temperatures stabilized by a capacitance sensor
with an accuracy of about 1 mK. The angular dependence of the MR was
determined by rotating the sample within a 100$^{\circ}$ range under
constant magnetic fields up to 16 T (Cernox sensor was used for
the temperature control during the rotation).

\begin{figure}[t]
\epsfxsize=8.5cm
\centerline{\epsffile{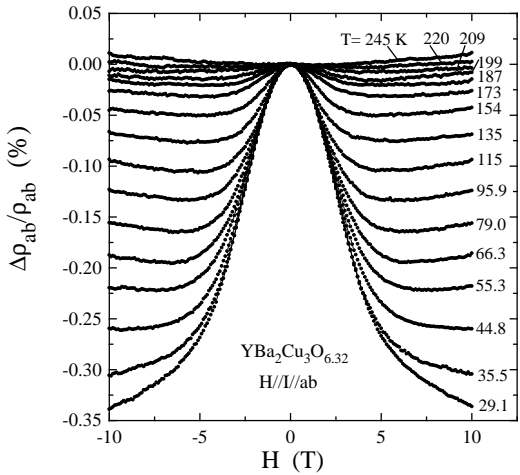}}
\vspace{2mm}
\caption{In-plane longitudinal MR of YBa$_2$Cu$_3$O$_{6.32}$.
The data are averaged over several field sweeps.}
\label{fig2}
\end{figure}

Figure 1 shows the in-plane resistivity $\rho_{ab}$ of the heavily
underdoped YBa$_2$Cu$_3$O$_{6+x}$ crystals studied in this work. One can
see here that the evolution of the in-plane transport upon reducing the
carrier concentration is much more gradual than is often considered; in
Fig. 1, apart from the low-$T$ region below $\sim$50 K, $\rho_{ab}(T)$
retains a metal-like behavior even in these samples with $x$$\simeq$0.30 or
0.32, which are located deep in the AF region of the phase diagram (see
inset to Fig. 1). The growth of $\rho_{ab}$ at low temperatures is notably
slower than $A\exp {(B/T)^k}$ with $k$ = 1/4 - 1, which also implies that
the system we are dealing with is not a simple band-gap insulator nor an
Anderson insulator.

These crystals demonstrate an unusual behavior of the in-plane MR,
$\Delta\rho_{ab}/\rho_{ab}$, when the magnetic field $H$ is applied along
the CuO$_2$ planes. Figure 2 shows the in-plane MR measured in the
longitudinal geometry [$H$$\parallel$$I$$\parallel$$a(b)$]. At low fields,
this longitudinal in-plane MR is negative and follows roughly a
$T$-independent parabolic curve $\zeta H^2$, but abruptly saturates above
some threshold field. The threshold field $H_{th}$ and the magnitude of the
saturated MR gradually increase with decreasing temperature. Above
$H_{th}$, the MR can be fitted with the usual $\gamma H^2$ dependence;
namely, we can write the $H$ dependence of the MR above $H_{th}$ as
$\Delta\rho_{ab}/\rho_{ab}$ = $(\Delta\rho_{ab}/\rho_{ab})_0$ + $\gamma
H^2$. Therefore, the behavior of the MR in Fig. 2 can be viewed as a
superposition of the low-field feature [whose size can be measured by
$(\Delta\rho_{ab}/\rho_{ab})_0$] onto a weak background MR of $\gamma H^2$.
Note that $\gamma$ is positive at high $T$ and changes its sign to negative
at about 50 K, where $\rho_{ab}(T)$ acquires the localizing behavior (see
Fig. 1).

\begin{figure}[t]
\vspace{-15pt}
\leftskip-10pt
\epsfxsize=1.0\columnwidth
\epsffile{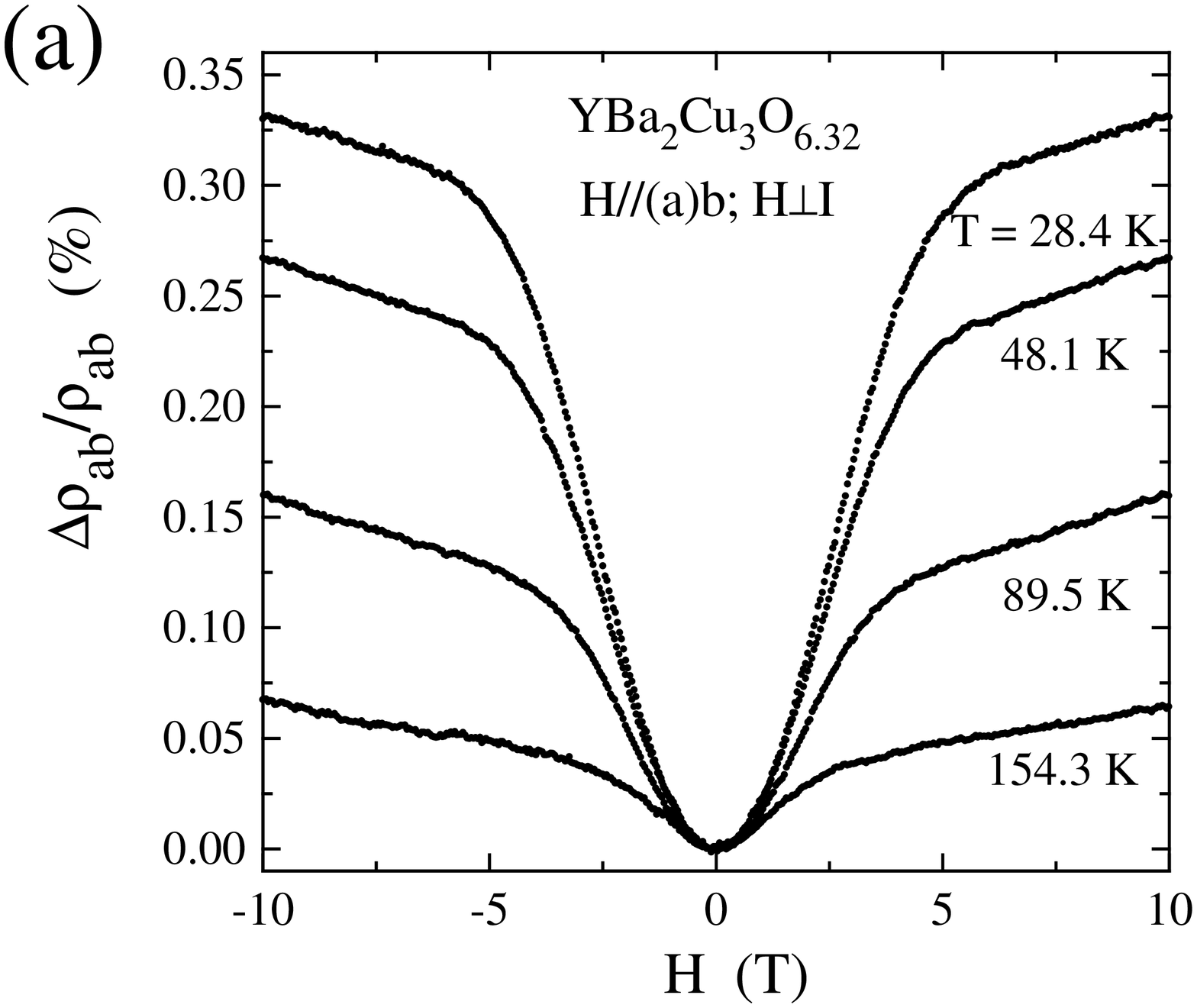}
\epsfxsize=1.2\columnwidth
\epsffile{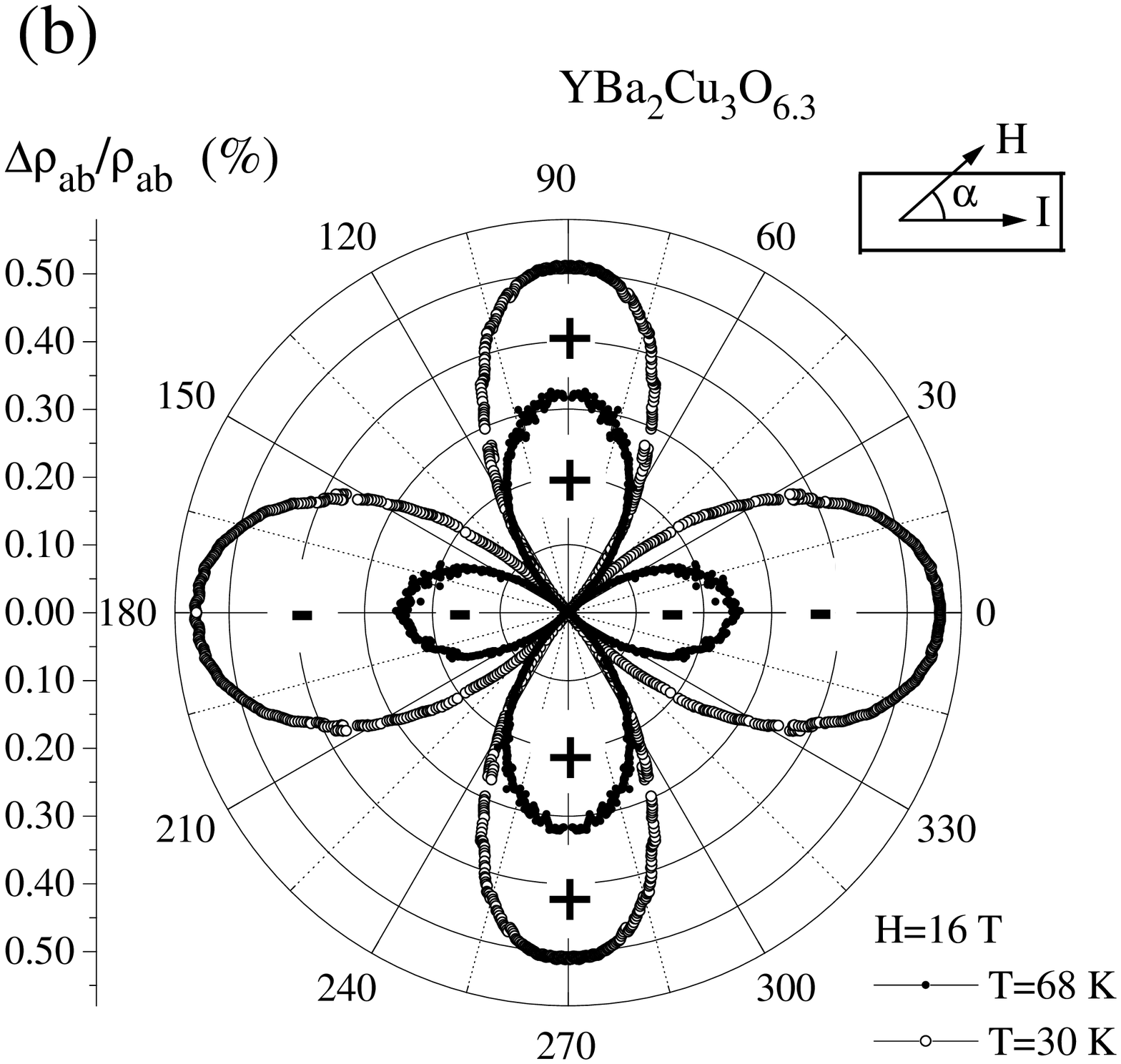}
\caption{(a) In-plane MR for the same sample as in
Fig.~\ref{fig2}, but with the field turned transverse to the current
[$H$$\parallel$$a(b)$, $H$$\perp$$I$].
(b) Angular dependences of the MR [$H$$\parallel$$(ab)$; $H$=16 T] for
YBa$_2$Cu$_3$O$_{6.3}$; the sign of MR is indicated.
Actual data were taken for two quadrants at 68 K and for one quadrant
at 30 K, and then extended to all angles.}
\label{fig3}
\end{figure}

To our surprise, when the magnetic field is turned in the plane and becomes
perpendicular to the current [$H$$\parallel$$a(b); H$$\perp$$I$], we find
that the low-field MR term just switches its sign, retaining its magnitude
and the threshold-field value [Fig. 3(a)]. We thus performed a detailed MR
measurements upon rotating $H$ within the $ab$ plane, which revealed a
striking anisotropy with a ``$d$-wave" like symmetry [Fig. 3(b)]; namely,
$\Delta\rho_{ab}/\rho_{ab}$ changes from negative at $\alpha$=$0^{\circ}$
($\alpha$ is the angle between $H$ and $I$) to positive at
$\alpha$=$90^{\circ}$, passing through zero at about $45^{\circ}$. Note
that the MR diagram in Fig. 3(b) is fairly symmetric. Some $T$-dependent
difference between the longitudinal and transverse segments is mainly
caused by the background MR; at 68 K, for example, 
the positive $\gamma H^2$ term extends
the ``$+$'' arm at $\alpha=90^{\circ}$ and reduces the ``$-$'' arm at
$\alpha=0^{\circ}$.

\begin{figure}[t]
\vspace{-20pt}
\epsfxsize=1.0\columnwidth
\epsffile{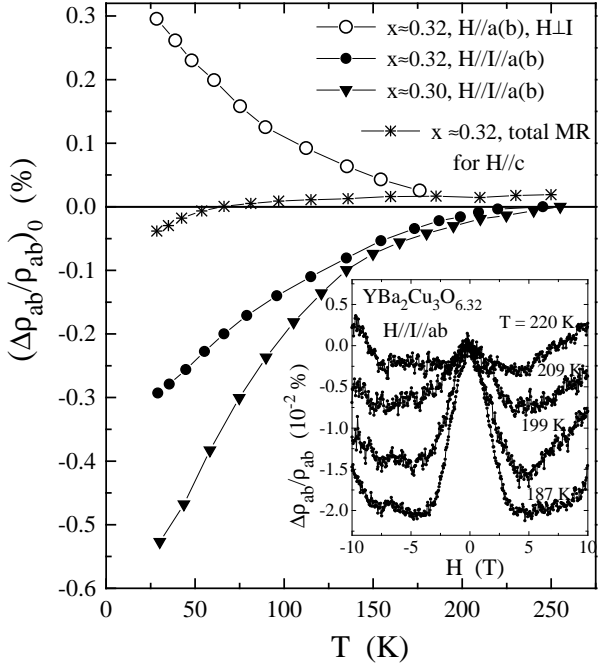}
\vspace{-36pt}
\caption{Magnitude of the low-field MR for $H$$\parallel$$I$ and
$H$$\perp$$I$.  For comparison, total magnitude of the MR at 10 T is
shown for $H$$\parallel$$c$, where there is no low-field anomaly.
Inset: $H$ dependences of the longitudinal MR at high $T$.}
\label{fig4}
\end{figure}

It is worth noting that this low-field MR feature is not observed at all
when the magnetic field is applied along the $c$-axis; the in-plane MR for
$H$$\parallel$$c$ is weak and approximately similar to the $\gamma H^2$
term in the longitudinal MR (see also Ref. \cite{ourMR}). This means that
the orbital part of the MR and interference effects are irrelevant to the
anomalous low-field behavior observed here. 

Since the samples being studied here are antiferromagnets, it is natural to
look for a connection between the long-range AF order and the observed
behavior. The N\'{e}el temperature $T_N$ for these two samples with
$x$=0.30 and 0.32 has been evaluated to be about 230 and 200 K,
respectively, from the measurements of $c$-axis resistivity $\rho_c$
\cite{ourMR} (the slope of $\rho_c(T)$ shows a distinct anomaly at $T_N$
\cite{Lavrov}).  The sharpness of the anomaly in $\rho_c$ tells us a very 
high homogeneity of the oxygen in our samples \cite{ourMR}. 
The magnitude of the low-field MR term,
$(\Delta\rho_{ab}/\rho_{ab})_0$, plotted in Fig. 4 as a function of
temperature, actually becomes noticeable near $T_N$ and
gradually grows with decreasing temperature. However, it is not clear
whether the long-range N\'{e}el order itself plays a dominant role in the
low-field MR anomalies, because $(\Delta\rho_{ab}/\rho_{ab})_0$ does not
show an abrupt increase at $T_N$ (which at least tells us that the feature
is not related to the spin-flop transition). It is more likely that the
anomaly is governed by the AF correlation within the CuO$_2$ planes,
rather than the long-range order. It is worth noting that even at high $T$
(near 200 K) the low-field MR feature remains sharp with a well-defined
threshold field (inset to Fig. 4). This suggests that the anomaly is a
cooperative phenomenon; if the anomaly is just an integration of the
behaviors of independent spins, the feature would be smeared by the thermal
fluctuations at such high temperatures.
We also note that, since Y has no magnetic moment, the anomaly observed 
here cannot be associated with magnetic ordering of the rare-earth 
sublattice.

\begin{figure}[t]
\epsfxsize=10cm
\centerline{\epsffile{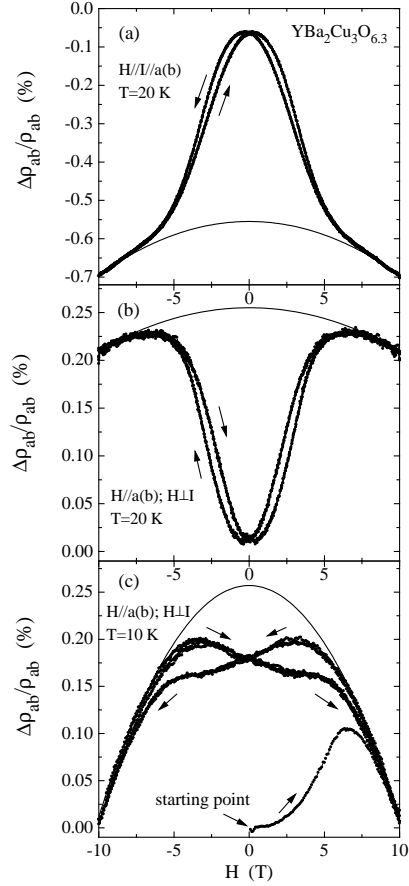}}
\vspace{2mm}
\caption{Hysteresis in the low-$T$ MR. Each curve contains data of
4 field sweeps (no averaging was made) performed at a rate of 1 T/min. In
panel (c), the first run is different from subsequent ones and the
resistivity does not return to its zero-field value. The solid lines
depicts the background MR.}
\label{fig5}
\end{figure}

The most intriguing peculiarity of the weak-field MR becomes evident at low
temperatures. As the temperature decreases below 20 - 25 K, the
$H$-dependence of $\rho_{ab}$ becomes irreversible. Initially this
irreversibility appears as a small hysteresis on the MR curve
[Figs. 5(a) and 5(b)].
However, as can be seen in Fig. 5(c), the irreversibility becomes much more
pronounced upon cooling to 10 K. In Fig. 5(c), $\rho_{ab}(H)$ curve similar
to that at higher temperatures can be observed only in the first sweep
which starts from $\Delta\rho_{ab}/\rho_{ab}$=0; the data from the
subsequent field sweeps significantly differ from the first one, with a
strongly reduced low-field feature [the peak hight is suppressed and the
position of the peak is further shifted compared to the peaks in Fig.
5(b)]. The important point here is that the resistivity does not return to
its initial value after removing the magnetic field; namely, the system
acquires a memory. The application of the magnetic field at low
temperatures therefore results in a persistent change in the resistivity;
it can be easily seen that the sign and the size of this persistent change
depends on the field direction (with respect to the current) according to
the diagram in Fig. 3(b). This indicates that the magnetic field induces a
persistent anisotropy of the in-plane resistivity, which has the
``$d$-wave" like symmetry of Fig. 3(b).

It is very difficult to understand the novel MR features found here,
especially the memory effect, without considering an inhomogeneous state or
a superstructure in the CuO$_2$ planes instead of a uniform AF state. The
observed hysteresis tells us that the structure is frozen in below $\sim$20
K, while it is mobile at high and moderate temperatures. There are only two
candidates which can produce a non-uniform structure: the ion rearrangement
and the hole segregation. Since one can hardly imagine some rearrangement
of ions to occur at such low temperatures and to be driven by the magnetic
field, the holes can be the only possible factor responsible for the
inhomogeneous structure.

The picture of charge inhomogeneities in the CuO$_2$ planes with holes
being confined to the ``stripes'' \cite{Emery} 
actually allows to account for all the observed MR
peculiarities. Since the AF interactions are essential for the formation of
the stripes, it is expected that the development of the stripes is closely
tied to the growth of AF domains in the CuO$_2$ planes. Indeed, for the
heavily underdoped region with the long-range AF order, the measurements of
the staggered magnetization and the N\'{e}el temperature \cite{AFstrip1}
have provided a substantial support for the presence of the striped phase
\cite{AFstrip2}. It is naturally expected that a strong local anisotropy is
introduced by the stripes with confined carriers moving along; however,
such an anisotropy will not be observed at long length scales if the
orientational order of the stripes is not established or the stripe
direction alters from one CuO$_2$ plane to another. Within this picture,
one can imagine that the magnetic field would give rise to a topological
ordering of the stripes, aligning them along the field direction and
changing the array of the current paths. The existence of the threshold
field for the low-field feature is presumably coming from the establishment
of the directional order of the stripes. Also, the ``$d$-wave" like 
anisotropy of the in-plane MR can be understood as a consequence of the 
rotation of the stripe direction with respect to the current direction.

The observation that the stripes can be directed with external magnetic
field implies that the stripes should have local ferromagnetic moment.
(Otherwise the stripes cannot couple to the magnetic field.) This is
possible if the stripes are ``in-phase boundaries", at which the spin
directions across the boundary are the same. 
We note that the MR behavior observed here 
surprisingly resembles that of a ferromagnetic metal with a mesoscopic
domain structure \cite{ferro}.

As the temperature is lowered, it is expected that the stripe dynamics
slows down and the magnetic domain structure in the CuO$_2$ planes is
frozen, forming a cluster spin glass \cite{glass}. The spin-glass
transition temperature has been reported for heavily-underdoped
Y$_{1-y}$Ca$_y$Ba$_2$Cu$_3$O$_6$ to be about 20 - 25 K for the AF
compositions \cite{glass}, which is in good agreement with the temperature
where the hysteretic MR behavior is found to show up in our experiment.

In summary, the in-plane magnetoresistance (MR) in heavily underdoped
YBa$_2$Cu$_3$O$_{6+x}$ is found to demonstrate a variety of unusual
features, including a striking ``$d$-wave" shaped angular dependence, 
anomalous low-field MR and its saturation above 
a well-defined threshold field, and a
hysteretic behavior at low temperatures. The overall features can be
consistently explained by assuming a developed array of charged stripes and
the field-induced topological ordering of the stripes. At temperatures below
$\sim$10 K, the external magnetic field of the order of a few T can produce
a persistent directional ordering of the stripes, giving rise to a ``memory
effect" in the resistivity. These findings give a strong evidence that the
charge inhomogeneities in the CuO$_2$ planes in the form of stripes exist
and actually have a considerable impact on the electron transport.
Also, our data show that the magnetic field
can be used as a tool to manipulate the striped structure.

A.N.L. gratefully acknowledges support from JISTEC.

%

\medskip
\vfil

\end{document}